\documentclass[12pt,a4paper]{panl}
\usepackage{cite}
\usepackage{wrapfig}
\usepackage{graphicx}
\usepackage{amssymb}
\usepackage{amsfonts}
\usepackage{amsmath}
\usepackage{longtable}
\usepackage{rotating}
\usepackage{lscape}
\usepackage{epsfig}
\usepackage{multirow}

\originalTeX
\begin{document}

\title{Search for dibaryon resonances  in the reactions  $dd\to dd\pi\pi$ and $pd \to pd\pi\pi$ \\
}
\maketitle
\authors{Yu.N.\,Uzikov$^{a,b,c}$\footnote{E-mail: uzikov@jinr.ru},
A.A.\,Temerbayev$^{d,e}$\footnote{E-mail: temerbayev\_aa@enu.kz}, 
N.T.\,Tursunbayev$^{d}$\footnote{E-mail:tursunbayev.n@gmail.com}}
\setcounter{footnote}{0}
%
\from{$^{a}$\,Dzhelepov Laboratory of Nuclear Problems, Joint Institute for Nuclear Researches, Dubna, Russia}
\from{$^{b}$\, Moscow State University, Physics Faculty, Moscow, Russia }
\from{$^{с}$\,$^{c}$\,Dubna State University, Dubna, Russia}
%
\from{$^{d}$\,Gumilyov Eurasian National University, Astana, Republic of Kazakhstan}
%
\from{$^{e}$\,Institute of Nuclear Physics, Astana branch, Astana, Republic of Kazakhstan} 

\begin{abstract}
%
The isoscalar dibaryon resonance $D_{03}(2380)$, discovered at WASA@COSY
in the total cross section of the  reaction  $pn\to d\pi^0\pi^0$ and in
the elastic pn scattering
in the energy range corresponding to the invariant mass of the pn system of 2380 MeV, can  reveal itself
in the $pd\to pd\pi\pi$ and $dd\to dd\pi\pi$ reactions at higher energies due to deuteron excitation
in the $t$ channel. In this paper, the cross sections of these reactions are estimated based on the
model of the reaction
$pn\to d\pi^0\pi^0$, proposed earlier by Platonova and Kukulin.
\end{abstract}
\vspace*{6pt}


\label{sec:intro}
\section*{Introduction}












\label{sec:Introduction}
In theory an idea of existence of dibaryon resonances appeared in 1964 almost simultaneously with the
 appearence of the quark picture of hadrons and some  predictions for masses and  quantum numbers of
 nonstrange dibaryons were done \cite{Dyson:1964xwa}.
In particular, the isoscalar $T=0$ resonance with spin-parity $J^P=3^+$  was predicted within  the SU(6)-symmetry
 with the mass 2350 MeV.
In experiment first indications  to possible effects connected with excitation of dibaryon resonances
appeared only in 1977 \cite{Kamae:1976as} and then in \cite{Ikeda:1978ika}, where the angular dependence of proton
polarization in the reaction $\gamma d\to pn$
 has been measured at several  hundred MeV of  photon energy. The polarization and differential-cross-section
 data on this reaction were consistently explained by introducing a dibaryon resonance
 $T(J^+) = 0(3^+)$ or $0(1^+)$ at  mass of  2360 MeV. At present one of the most realistic candidate to
 dibaryon resonance  is the  resonance $D_{IJ}=D_{03}$  observed  by the WASA@COSY
 \cite{WASA-at-COSY:2011bjg,WASA-at-COSY:2012seb}
 in the total cross section  of the reaction $pn\to d\pi^0\pi^0$ with the isospin $I=0$ and spin $J=3^+$.
 Mass of the resonance is 2380 MeV that is close to the $\Delta\Delta$ or $NN^*(1440)$  thresholds,
  however the width is 70 MeV that is   several times smaller than the width of two  free $\Delta$-s or one $N^*(1440)$.
  This narrow width is the most clear  indication to  a non-usual hadron structure of the observed resonance state.
   In theory, there are considered  quasi-molecular  $\pi N\Delta$  state \cite{Gal:2013dca}
  or $\Delta\Delta$  plus   quark component with hidden color \cite{Dong:2016rva}.
  The excitation of this resonance can be related to the  well known ABC effect  observed
 in  $pd\to ^3He \pi\pi$ reactions \cite{Abashian:1960zz}
  with
  two meson production in  isoscalar state and not explained in theory.
  A model of the reaction $pn\to d\pi^0\pi^0$ based on the  sum of the
  mechanisms with the $D_{03}\to\sigma +d$ decay  and two resonance decay  $D_{03}\to D_{12}+\pi$, $D_{12}\to \pi+ d$
 was suggested  in  Ref. \cite{Platonova:2012am}. Later on this model was modified by incluson of the  mechanism with the
 $D_{03}\to \Delta\Delta$ decay \cite{Platonova:2021trw}.
 An observation of this resonance in the reaction $pd\to pd \pi\pi$  was made by ANKE@COSY \cite{Komarov:2018pry}
 at kinetic energies of proton beam 1-2 GeV.   An  attempt to explain these data on the basis of the properly modified
 model \cite{Platonova:2012am} of the reaction  $pn\to d\pi^0\pi^0$ assuming t-channel excitation of the deuteron
 into the dibaryon
 $D_{03}$ via the $\sigma$-meson exchange was done  in Refs.\cite{Uzikov:2019vey}, \cite{Tursunbayev:2020dif}

  Reviews
 of works  devoted to the $D_{03}$  resonance were done
  in Refs. \cite{Clement:2016vnl}  and \cite{Clement:2020mab}.
   Indications to the $D_{03}$ resonance and  more heavy isoscalar resonances with masses  of  2470 MeV and 2630 MeV
   and  width  $\sim 120$ MeV were found in the reaction $\gamma + d\to d+\pi^0\pi^0$
  by the  ELPH collaboration in \cite{Ishikawa:2018wkv} and recently  by the
BGOOD  collaboration \cite{Jude:2022atd}
  .

\begin{figure}[t]
\begin{center}
\includegraphics[width=55mm]{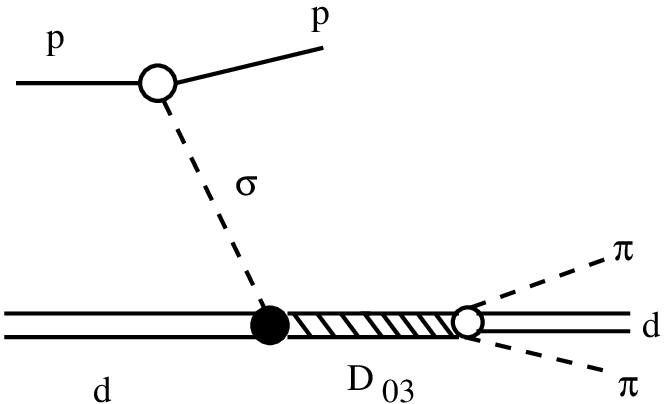}
\includegraphics[width=50mm]{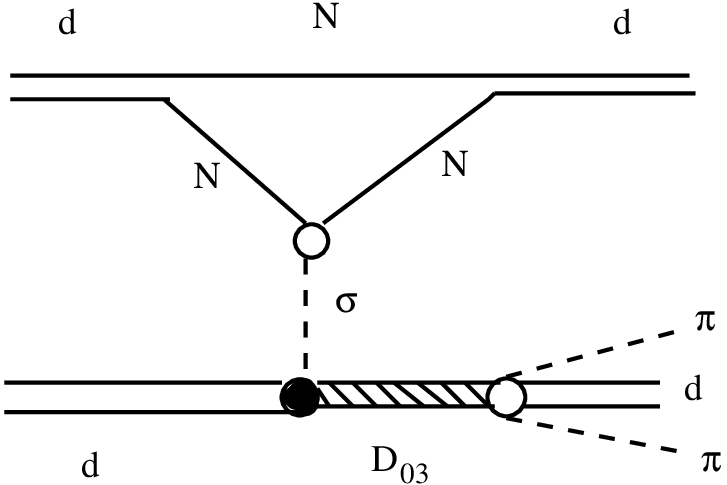}
\vspace{-3mm}
\caption{The  mechanism of the deuteron excitation  via the $\sigma$-meson exchange
in the reactions
$pd\to pd\pi\pi$ (left)
 and $dd\to dd\pi\pi$ (right).
}
\end{center}
\labelf{fig-mech1}
\vspace{-5mm}
\end{figure}

     In present work we consider a possible search for
     the $D_{03}(2380)$ resonance and also  more heavy isoscalar resonances, discussed  in
     Refs.\cite{Ishikawa:2018wkv,Jude:2022atd},
    in the reactions  $dd\to dd\pi\pi$ and $pd \to pd\pi\pi$
     at SPD NICA \cite{SPD:2024gkq}.
     Since collision energy at SPD will be rather high, $\sqrt{s_{dd}}\leq 8 $ GeV, the excitation of these
     dibaryons in the s-channel of pn-collison is impossible. However the excitation via the  t-channel exchanges,
      as it was shown by ANKE
      \cite{Komarov:2018pry}, is possible in  the   $pd\to pd\pi\pi$  and, probably, in the $dd\to dd\pi\pi$ reactions.

        In the next section we give an estimation of the cross section of the reaction   $pd\to pd\pi\pi$ at SPD NICA  energies
        in the line of works Refs. \cite{Uzikov:2019vey,Tursunbayev:2020dif}
         assuming that this is a subprocess of the reaction $dd\to n+pd\pi\pi$ considered  in the impulse approximation.
        After that  the cross section of the  reaction $dd\to dd\pi\pi$ is estimated
        using a properly modified mechanism of the $pd\to pd\pi\pi$ reaction  with the excitation of the  $D_{03}$ resonance. Conclusion is done in the  last section.


\section{Differential cross sections of the   reactions $pd \to pd\pi\pi$ and  $dd\to dd\pi\pi$}
\subsection{The  reaction $pd \to pd\pi\pi$.}

 \,\, In calculation of the cross sections of the   reaction $pd \to pd\pi\pi$ we use  the  mechanism of excitation of the
 deuteron to the $D_{03}$ resonance  by exchange of the  $\sigma$-meson  between the proton and deuteron (Fig. \ref{fig-mech1},
 left). According to the model of Ref. \cite{Platonova:2012am}, forthcoming decay of the $D_{03}$  resonance
  includes sum of two mechanisms: (i) the  $D_{03}\to \sigma+ d$ decay, which is followed by the  $\sigma\to \pi\pi$ decay
  and  (ii) two  sequential decays $D_{03}\to \pi_1+ D_{12}$ and  $D_{12}\to \pi_2+d$. The necessary formalism is given in Ref.\cite{Uzikov:2019vey}.
  The differential cross section for distribution on the invariant mass of the final $\pi\pi d$ system,  $M_{d\pi\pi}$,
  has the following form
  \begin{eqnarray}
  \label{difcx}
 \frac{d\sigma}{dM_{d\pi\pi}}=\frac {C_T}{64(2\pi)^8 |{\bf p}_f| s}\int dM_{\pi\pi}
 d\Omega_{{\bf p}_f}
 d\Omega_{\bf k} d\Omega_{\bf q} |{\bf k}|\,|{\bf q}|\,|{\bf p}_f|
  \overline{|M_{fi}|^2};
  \end{eqnarray}
 where
 $M_{\pi\pi}$ is the invariant mass of the $\pi\pi$ system, $s$ is the pd-invariant mass squared,
 ${\bf p}_i$  (${\bf p}_f$) is the 3-momentum of the  initial (final) proton  in the  c.m. system of the reaction
  $pd \to pd\pi\pi$,
${\bf k}$ is the relative momentum in the $\pi\pi$ system, ${\bf q}$ is the deuteron momentum in the $\pi\pi d$ c.m. system,
  $C_T$ is the isopsin factor. The  $M_{fi}$ in Eq. (\ref{difcx}) is the matrix element of the reaction  $pd \to pd\pi\pi$,
  which  has the following form for the  $\sigma$-exchange mechanism  shown in Fig. \ref{fig-mech1} (left)
\begin{eqnarray}
\label{amp-pd}
M_{fi}= M(p\to p'\sigma)\frac{1}{p_\sigma^2-m_\sigma^2+im_\sigma \Gamma_\sigma} M(\sigma d\to d\pi\pi),
\end{eqnarray}
here  $p_\sigma$ is the 4-momentum of the $\sigma$-meson, $m_\sigma$ and $\Gamma_\sigma$ are its mass and width.
The vertex $\sigma pp$ amplitude $M(p\to p'\sigma)$ is taken from Ref.
\cite{Cao:2010km}
and its squared spin averaged form
is the following
\begin{eqnarray}
\overline{ |M(p\to p'\sigma|^2}=\Bigl (\frac{f_\sigma}{m_\sigma}\Bigr )^2 (2m)^2
\Bigl [D-2{\bf p}_i{\bf p}_f+\frac{(|{\bf p}_i| |{\bf p}_f|)^2}{D}\Bigr ] F_{\sigma }(p_\sigma^2,\lambda_\sigma)^2,
\end{eqnarray}
\begin{figure}[t]
\begin{center}
\includegraphics[width=77mm]{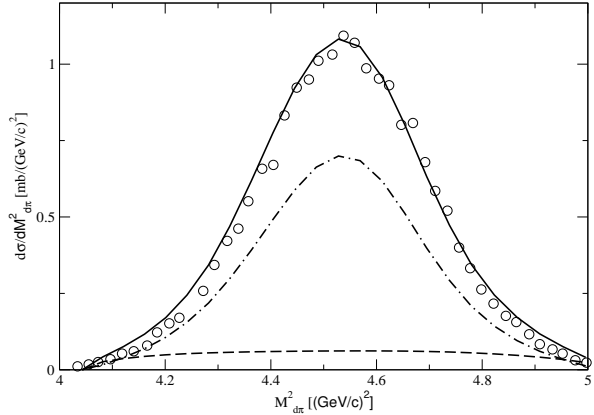}
\vspace{-3mm}
\caption{The differential cross section as a function  of the invariant mass squared  $M^2_{d\pi}$
in the reaction $pn \to d \pi^0\pi^0$ at the total energy
$\sqrt{s_{pn}}=2.380$ GeV.
The contribution of the $\sigma$-meson production mechanism is shown by dashed line, while the mechanism with intermediate
 dibaryon $D_{12}$ is  shown by dashed-dotted line and the full line corresponds to the sum of $\sigma$ and $D_{12}$ mechanisms.
 The experimental data (circles) are taken from  Ref.\cite{WASA-at-COSY:2011bjg}.}
\end{center}
\labelf{fig00}
\vspace{-5mm}
\end{figure}

where $D=(E_i+m)(E_f+m)$ and $m$ is the nucleon mass, $E_i$ ($E_f$) is the energy of the initial (final) proton
in the c.m. system of the reaction $pd \to pd\pi\pi$;
$F_{\sigma}(p_\sigma^2,\lambda_\sigma)^2=(\lambda_\sigma^2 -m_\sigma^2)/(\lambda_\sigma^2 -p_\sigma^2)$.
The amplitude $M(\sigma d\to d\pi\pi)$  is given in Ref.  \cite{Uzikov:2019vey} in the framework of the model  \cite{Platonova:2012am}  for the
reaction $pn\to d \pi^0\pi^0$  and contains  the amplitude $\sigma d\to D_{03}$ in product with i)
 the $D_{03}\to \sigma d$ and $\sigma \to \pi\pi$ amplitudes for the $\sigma$-mechanism and ii)
 the $D_{03}\to D_{12}\pi$  and $D_{12}\to \pi d$ amplitudes for the $D_{12}$ mechanism with the dibaryon
 $D_{12}$ which  isospin  is $T=1$, angular momentum $J=2$ and mass  $2150$ MeV. All these vertex amplitudes correspond
  to the decay channel of the type  $R\to a+b$  and contain
   the vertex factor $F_{R\to ab}$  defined
  in Ref.\cite{Platonova:2012am} as
  \begin{eqnarray}
  \label{FRab}
  F_{R\to ab} =M_{ab} \sqrt{\frac {8\pi\Gamma^l_{R\to ab}(q_{ab})} {(q_{ab})^{2l+1}}}, \\ \nonumber
  \Gamma^l_{R\to ab}(q_{ab})=\Gamma^l_{R\to ab}(q_0)\Bigl(\frac{q}{q_0}\Bigr )^{2l+1}
  \Bigl(\frac{q_0^2+\lambda_{ab}^2}{q^2+\lambda_{ab}^2} \Bigr )^{l+1},
  \end{eqnarray}
where $M_{ab}$ is the mass  of the system $a+b$, $q_{ab}$ is the relative momentum  in the $a+b$  c.m. system, $q_0$ is  the corresponding momentum at the resonance point $M_{ab}=M_R$, $l$ is the orbital momentum in the decay $R\to a+b$;
$\lambda_{ab}$ is the parameter of the vertex form factor, $\Gamma^l_{R\to ab}(q_0)\equiv \Gamma^l_{R\to ab}$ is the total partial width
in the decay channel $R\to a+b$. The orbital momenta for considered vertices are $l=2$ for $D_{03}\leftrightarrow d\sigma$,
 and $l=1$  for the  both decays $D_{03}\to D_{12}\pi$ and $D_{12}\to d\pi$.
 Contribution of other mesons $X$ into this mechanism  is not considered here, since their coupling to the $X d^*d$ vertex is unknown.


 \subsection{The  reaction $dd \to dd\pi\pi$.}
 \,\,
  The mechanism of the reaction  $dd \to dd\pi\pi$  includes the elastic  deuteron form factor  $S_d(\Delta/2)$ in the upper vertex
  in Fig. \ref{fig-mech1} (right).
   Therefore, the corresponding transition amplitude $M_{fi}(dd\to dd\pi\pi)$ can be obtained
  from   Eq.(\ref{amp-pd}) by   multiplying  it by the form factor  $S_d(\Delta/2)$,
   where  the transferred  3-momentum is $\Delta=\sqrt{(-p_\sigma^2)}$.
  In view of isoscalar nature of the $\sigma$-meson and the deuteron, the upper vertex in Fig. \ref{fig-mech1} (right) does not contain additional isospin
  factor as compared to Eq. (\ref{difcx}).
\begin{figure}[t]
\begin{center}
\includegraphics[width=87mm]{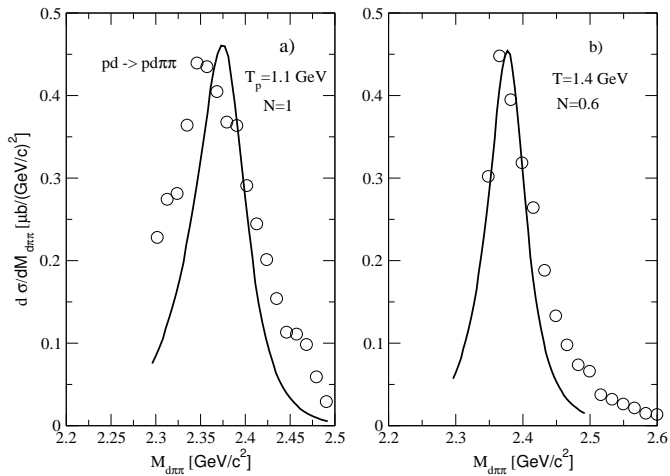}
\vspace{-3mm}
\caption{The distributions over the invariant mass $M_{d\pi\pi}$ in the reaction $pd\to pd \pi\pi$ at the proton beam kinetic energy
 1.1 GeV ({\it a}) and 1.4 GeV ({\it b}). The results of calculation (full curves) according to mechanism in Fig. \protect{\ref{fig-mech1}} (left panel)
 in comparison with experimental data ($\circ$) from  Ref.\cite{Komarov:2018pry}.}
\end{center}
\labelf{fig01}
\vspace{-5mm}
\end{figure}

 \section{Numerical results and discussion}

  In numerical calculations we use  the following numbers for masses and width of the resonances $D_{03}$ and $D_{12}$  and the vertices parameters: $M_{D_{03}}=2.380 $ GeV, $\Gamma_{D_{03}}= 70$ MeV, $M_{D_{12}}=2.15 $ GeV, $\Gamma_{D_{12}}=0.11$ GeV
 $f_\sigma=2.20$, $m_\sigma=0.55$ GeV, $\Gamma_\sigma=0.5$ GeV,
 $\lambda_\sigma=1.3$ GeV \cite{Cao:2010km}. The values $\Gamma^{(l=1)}_{D_{12}\to d\pi}=10 $ MeV,
 $\lambda_{d\sigma}=0.18$ GeV,
 $\lambda_{d\pi}=0.25$ GeV obtained in Ref. \cite{Platonova:2012am} are used here.
 Partial widths  $\Gamma^{(2)}_{D_{03}\to d\sigma}$ and $\Gamma^{(2)}_{D_{03}\to D_{12}\pi}$ were  not determined
 in Ref.\cite{Platonova:2012am}. Later on a possible solution  for these  parameters was  found in Ref. \cite{Platonova:2021trw}, but an addtional mechanism  with the decay channel $D_{03}\to \Delta\Delta$ was included. We are not able to reproduce the numerical results of the paper \cite{Platonova:2021trw} and, therefore, restrict ourself by the first model developed in Ref. \cite{Platonova:2012am}. We assume here $\Gamma^{(2)}_{D_{03}\to d\sigma}=20$ MeV and $\Gamma^{(1)}_{D_{03}\to D_{12}\pi}=90$ MeV,
  because with these parameters the experimental data on differential cross section $d\sigma/dM_{d\pi}$ of the reaction $pn\to d\pi^0\pi^0$  \cite{WASA-at-COSY:2011bjg}  are well reproduced in  present  calculations (see Fig. \ref{fig00}) using  the model \cite{Platonova:2012am}.
  In these calculations we use the sum of two mechanisms for the $D_{03}\to d\pi\pi$ decay, i.e.
  $D_{03}\to d\sigma \to d \pi\pi$ and   $D_{03}\to D_{12}\pi \to d\pi\pi$ with above written parameters, and include
  the fusion amplitude  $pn\to D_{03}$  in the form of Eq.(\ref{FRab})  with parameters taken  from Ref.\cite{Platonova:2012am}.
  The isospin factor $C_T$ accounting for pairs of charged and neutral pions is $C_T=2.8$ \cite{Uzikov:2019vey}.
  The intervals of the deuteron scattering  angles in the reaction $dd\to dd\pi\pi$ are $8^\circ-13^\circ$ for the "forward"  \cite{SPD:2024gkq}
  and $172^\circ-167^\circ$ for the "backward" direction.

\begin{figure}[t]
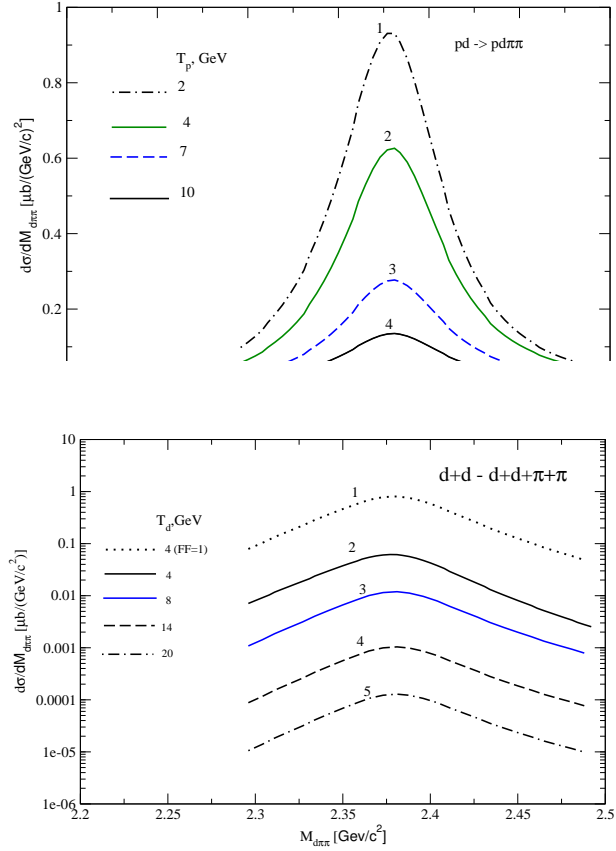

\begin{center}
\includegraphics[width=80mm]{pdpd-2-10m.eps}
\includegraphics[width=80mm]{dd2pi-FF.eps}
\vspace{-3mm}
\caption{The distributions over the invariant mass $M_{d\pi\pi}$ in the reaction $pd\to pd \pi\pi$ (upper panel)
and $dd\to dd \pi\pi$ (down panel)
calculated according to the mechanisms in Fig.\ref{fig-mech1}
at different
 proton beam  energy $T_p$ in $pd$- and  deuteron   beam   energy $T_d$  in dd-collision (at the total energy $\sqrt{s_{dd}}$):
(i) 1 -- $T_p =$ 2 GeV, 2 -- 4 GeV, 3 -- 7 GeV,  4 -- 10 GeV in the upper panel;
  (ii)  1, 2 --    $T_d=4$ GeV ( $\sqrt{s_{dd}}$= 5.4 GeV) , 3 -- 8 GeV ($\sqrt{s_{dd}}$= 6.64 GeV),  4 -- 14 GeV ($\sqrt{s_{dd}}$= 8.2 GeV),
 5 -- 20 GeV ($\sqrt{s_{dd}}$= 9.44 GeV), in the down panel. All lines in the down panel are obtained with the deuteron form factor included,
 except the line 1, where  the form factor is excluded.}
%
\end{center}
\vspace{-5mm}
\labelf{fig04}
\end{figure}
Results of our calculations of the differential  cross secton   $d\sigma/dM_{d\pi\pi}$ of the reaction $pd \to pd\pi\pi$
   in comparison with the experimental data  \cite{Komarov:2018pry} at proton beam energies $T_p=1.1$ and 1.4 GeV are shown in
    Fig. \ref{fig01}.  One can see that the theoretical curve  well reproduce the position of the resonance  peak  and its absolute
     magnitude  at 1.1 GeV, although at 1.4 GeV it overestimates the data by  factor of ~1.7.
     The results of calculations  at  higher energies 2-10 GeV are shown in Fig. \ref{fig04} (upper panel).
     One can see that with increasing energy the cross section  decreases quickly, that is caused to some extent by the factor $|{\bf p}_f|s$ in the denominator in
     Eq.(\ref{difcx}).  The $M_{d\pi\pi}$ distribution of the differential cross section of the reaction $dd \to dd\pi\pi$
     is shown in Fig. \ref{fig04} (down panel)  for  different energies $\sqrt{s_{dd}}=  5.4 - 9.4$ GeV.
     We have found that due to presence of the deuteron elastic form factor
  in the  transition amplitude
     of this process  the cross section is diminished  by two orders of magnitude as compared to the  corresponding cross section
    of the reaction  $pd \to pd\pi\pi$ given by Eq. (\ref{difcx}).

    \section{Conclusion}

    The differential cross sections of the reactions $pd \to pd\pi\pi$ and  $dd \to dd\pi\pi$ are calculated at  energies
   of SPD NICA  assuming t-channel excitation of the deuteron to the $D_{03}$ via the $\sigma$-meson exchange  between
    the beam particle  and  target deuteron with subsequent decay of the  dibaryon  $D_{03}$  to the $d\pi\pi$  system
    accoding to the model \cite{Platonova:2012am}. We do not try to make estimation for other more heavy  isoscalar dibaryons,
    since their structure and decay modes are unknown.
    Due to isospin conservation, in the reaction  $dd \to dd\pi\pi$ the final $\pi\pi$ system  can be only in the isoscalar state.
    Therefore this reaction constitutes  a clean test for  search  of isoscalar dibaryons.
    However, due to elastic  deuteron form factor the cross section  of this reaction is suppressed by one-two orders of magnitude as compared
     to the reaction $pd \to pd\pi\pi$. Nevertheless, the latter reaction can be studied as a subprocess in the reaction $dd\to n+p+d+\pi+\pi$
      in impulse approximation. Detection of the  $\pi^0\pi^0$ pair in the final state of the reaction $pd \to pd\pi\pi$
     selects  its  isoscalar channel and, therefore, an isoscalar dibaryon.
    If  two final mesons are not detected in this reaction, the isocalar dibaryon  can be  also visible in the  distribution
  over the $M_{d\pi\pi}$, as it was shown by  the ANKE data \cite{Komarov:2018pry}.

   {\bf FUNDING}.  This work is supported in part by the grant of the Scientific Programme of the Republic of
   Kazakhstan—JINR for
the 2025 year.


\bibliographystyle{pepan}
\bibliography{pepan_biblio}

\begin{thebibliography}{10}
\def\selectlanguageifdefined#1{
\expandafter\ifx\csname date#1\endcsname\relax
\else\selectlanguage{#1}\fi}
\providecommand*{\href}[2]{{\small #2}}
\providecommand*{\url}[1]{{\small #1}}
\providecommand*{\BibUrl}[1]{\url{#1}}
\providecommand{\BibAnnote}[1]{}
\providecommand*{\BibEmph}[1]{\emph{#1}}
\ProvideTextCommandDefault{\cyrdash}{\hbox to.8em{--\hss--}}
\providecommand*{\BibDash}{}

\bibitem{Dyson:1964xwa}
\selectlanguageifdefined{english}
\BibEmph{Dyson F., Xuong N.H.} {Y=2 States in Su(6) Theory}~//
  \href{http://dx.doi.org/10.1103/PhysRevLett.13.815}{Phys. Rev. Lett.}
  \BibDash
\newblock 1964. \BibDash
\newblock V.~13, no.~26. \BibDash
\newblock P.~815--817.

\bibitem{Kamae:1976as}
\selectlanguageifdefined{english}
\BibEmph{Kamae T., Arai I., Fujii T., Ikeda H., Kajiura N., Kawabata S.,
  Nakamura K., Ogawa K., Takeda T., Watase Y.} {Observation of an Anomalous
  Structure in Proton Polarization from Deuteron Photodisintegration}~//
  \href{http://dx.doi.org/10.1103/PhysRevLett.38.468}{Phys. Rev. Lett.}
  \BibDash
\newblock 1977. \BibDash
\newblock V.~38. \BibDash
\newblock P.~468.

\bibitem{Ikeda:1978ika}
\selectlanguageifdefined{english}
\BibEmph{Ikeda H., others.} {Angular Dependence of Proton Polarization in
  $\gamma d \to pn$ and Further Investigation of the Dibaryon Resonance}~//
  \href{http://dx.doi.org/10.1103/PhysRevLett.42.1321}{Phys. Rev. Lett.}
  \BibDash
\newblock 1979. \BibDash
\newblock V.~42. \BibDash
\newblock P.~1321.

\bibitem{WASA-at-COSY:2011bjg}
\selectlanguageifdefined{english}
\BibEmph{Adlarson P. et~al.} [WASA-at-COSY Collaboration] {ABC Effect in Basic
  Double-Pionic Fusion --- Observation of a new resonance?}~//
  \href{http://dx.doi.org/10.1103/PhysRevLett.106.242302}{Phys. Rev. Lett.}
  \BibDash
\newblock 2011. \BibDash
\newblock V. 106. \BibDash
\newblock P.~242302. \BibDash
\newblock arXiv:1104.0123~[nucl-ex].

\bibitem{WASA-at-COSY:2012seb}
\selectlanguageifdefined{english}
\BibEmph{Adlarson P. et~al.} [WASA-at-COSY Collaboration] {Isospin
  Decomposition of the Basic Double-Pionic Fusion in the Region of the ABC
  Effect}~// \href{http://dx.doi.org/10.1016/j.physletb.2013.03.019}{Phys.
  Lett. B}. \BibDash
\newblock 2013. \BibDash
\newblock V. 721. \BibDash
\newblock P.~229--236. \BibDash
\newblock arXiv:1212.2881~[nucl-ex].

\bibitem{Gal:2013dca}
\selectlanguageifdefined{english}
\BibEmph{Gal A., Garcilazo H.} {Three-Body Calculation of the Delta-Delta
  Dibaryon Candidate D(03) at 2.37 GeV}~//
  \href{http://dx.doi.org/10.1103/PhysRevLett.111.172301}{Phys. Rev. Lett.}
  \BibDash
\newblock 2013. \BibDash
\newblock V. 111. \BibDash
\newblock P.~172301. \BibDash
\newblock arXiv:1308.2112~[nucl-th].

\bibitem{Dong:2016rva}
\selectlanguageifdefined{english}
\BibEmph{Dong Y., Huang F., Shen P., Zhang Z.} {Decay width of $d^*(2380)\to NN
  \pi\pi$ processes}~//
  \href{http://dx.doi.org/10.1103/PhysRevC.94.014003}{Phys. Rev. C}. \BibDash
\newblock 2016. \BibDash
\newblock V.~94, no.~1. \BibDash
\newblock P.~014003. \BibDash
\newblock arXiv:1603.08748.

\bibitem{Abashian:1960zz}
\selectlanguageifdefined{english}
\BibEmph{Abashian A., Booth N.E., Crowe K.M.} {Possible Anomaly in Meson
  Production in p+d Collisions}~//
  \href{http://dx.doi.org/10.1103/PhysRevLett.5.258}{Phys. Rev. Lett.} \BibDash
\newblock 1960. \BibDash
\newblock V.~5. \BibDash
\newblock P.~258--260.

\bibitem{Platonova:2012am}
\selectlanguageifdefined{english}
\BibEmph{Platonova M.N., Kukulin V.I.} {ABC-effect as a signal of chiral
  symmetry restoration in hadronic collisions}~//
  \href{http://dx.doi.org/10.1103/PhysRevC.87.025202}{Phys. Rev. C}. \BibDash
\newblock 2013. \BibDash
\newblock V.~87, no.~2. \BibDash
\newblock P.~025202. \BibDash
\newblock arXiv:1211.0444~[nucl-th].

\bibitem{Platonova:2021trw}
\selectlanguageifdefined{english}
\BibEmph{Platonova M.N., Kukulin V.I.} {Isospin symmetry breaking in
  double-pion production in the region of d*(2380) and the scalar
  {\ensuremath{\sigma}} meson}~//
  \href{http://dx.doi.org/10.1103/PhysRevD.103.114025}{Phys. Rev. D}. \BibDash
\newblock 2021. \BibDash
\newblock V. 103, no.~11. \BibDash
\newblock P.~114025. \BibDash
\newblock arXiv:2105.12482.

\bibitem{Komarov:2018pry}
\selectlanguageifdefined{english}
\BibEmph{Komarov V.I., others.} {Resonance-like coherent production of a pion
  pair in the reaction $pd \rightarrow pd \pi\pi$ in the GeV region}~//
  \href{http://dx.doi.org/10.1140/epja/i2018-12641-0}{Eur. Phys. J. A}.
  \BibDash
\newblock 2018. \BibDash
\newblock V.~54, no.~11. \BibDash
\newblock P.~206. \BibDash
\newblock arXiv:1805.01493.

\bibitem{Uzikov:2019vey}
\selectlanguageifdefined{english}
\BibEmph{Uzikov Y., Tursunbayev N.} {Reaction of two pion production $pd \to
  pd\pi\pi$ in the resonance region}~//
  \href{http://dx.doi.org/10.1051/epjconf/201920408010}{EPJ Web Conf.} \BibDash
\newblock 2019. \BibDash
\newblock V. 204. \BibDash
\newblock P.~08010.

\bibitem{Tursunbayev:2020dif}
\selectlanguageifdefined{english}
\BibEmph{Tursunbayev N., Uzikov Y.} {The $D_{03}(2380)$ dibaryon resonance
  excitation in the $pd\to pd\pi\pi$ reaction.}~//
  \href{http://dx.doi.org/10.21468/SciPostPhysProc.3.056}{SciPost Phys. Proc.}
  \BibDash
\newblock 2020. \BibDash
\newblock V.~3. \BibDash
\newblock P.~056.

\bibitem{Clement:2016vnl}
\selectlanguageifdefined{english}
\BibEmph{Clement H.} {On the History of Dibaryons and their Final
  Observation}~// \href{http://dx.doi.org/10.1016/j.ppnp.2016.12.004}{Prog.
  Part. Nucl. Phys.} \BibDash
\newblock 2017. \BibDash
\newblock V.~93. \BibDash
\newblock P.~195. \BibDash
\newblock arXiv:1610.05591.

\bibitem{Clement:2020mab}
\selectlanguageifdefined{english}
\BibEmph{Clement H., Skorodko T.} {Dibaryons: Molecular versus Compact
  Hexaquarks}~// \href{http://dx.doi.org/10.1088/1674-1137/abcd8e}{Chin. Phys.
  C}. \BibDash
\newblock 2021. \BibDash
\newblock V.~45, no.~2. \BibDash
\newblock P.~022001. \BibDash
\newblock arXiv:2008.07200.

\bibitem{Ishikawa:2018wkv}
\selectlanguageifdefined{english}
\BibEmph{Ishikawa T., others.} {Non-strange dibaryons studied in the $\gamma
  d\to \pi^0\pi^0 d$ reaction}~//
  \href{http://dx.doi.org/10.1016/j.physletb.2018.12.050}{Phys. Lett. B}.
  \BibDash
\newblock 2019. \BibDash
\newblock V. 789. \BibDash
\newblock P.~413--418. \BibDash
\newblock arXiv:1805.08928.

\bibitem{Jude:2022atd}
\selectlanguageifdefined{english}
\BibEmph{Jude T.C., others.} {Evidence of a dibaryon spectrum in coherent
  {\ensuremath{\pi}}0{\ensuremath{\pi}}0d photoproduction at forward deuteron
  angles}~// \href{http://dx.doi.org/10.1016/j.physletb.2022.137277}{Phys.
  Lett. B}. \BibDash
\newblock 2022. \BibDash
\newblock V. 832. \BibDash
\newblock P.~137277. \BibDash
\newblock arXiv:2202.08594.

\bibitem{SPD:2024gkq}
\selectlanguageifdefined{english}
\BibEmph{Abazov V. et~al.} [SPD Collaboration] {Technical Design Report of the
  Spin Physics Detector at NICA}~// Natural Sci. Rev. \BibDash
\newblock 2024. \BibDash
\newblock V.~1. \BibDash
\newblock P.~1. \BibDash
\newblock arXiv:2404.08317.

\bibitem{Cao:2010km}
\selectlanguageifdefined{english}
\BibEmph{Cao X., Zou B.S., Xu H.S.} {Phenomenological analysis of the double
  pion production in nucleon-nucleon collisions up to 2.2 GeV}~//
  \href{http://dx.doi.org/10.1103/PhysRevC.81.065201}{Phys. Rev. C}. \BibDash
\newblock 2010. \BibDash
\newblock V.~81. \BibDash
\newblock P.~065201. \BibDash
\newblock arXiv:1004.0140~[nucl-th].

\end{thebibliography}
\end{document}